# Blind deblurring for microscopic pathology images using deep learning networks

**CHENG JIANG,**[1,†] **JUN LIAO,**[1,†] **PEI DONG,**[1] **ZHAOXUAN MA,**[1] **DE CAI,**[1] **GUOAN ZHENG,**[2] **YUEPING LIU,**[3] **HONG BU,**[4,5] **JIANHUA YAO**[1,*]

[1] *Tencent AI Lab, Shenzhen 518054, China.*
[2] *Department of Biomedical Engineering, University of Connecticut, Storrs, CT 06269, USA*
[3] *Department of Pathology, The Fourth Hospital of Hebei Medical University, Hebei 050011, China.*
[4] *Department of Pathology, West China Hospital, Sichuan University, Chengdu 610041, China.*
[5] *Laboratory of Pathology, Clinical Research Centre for Breast, West China Hospital, Sichuan University, Chengdu 610041, China.*
[†]*The two authors contribute equally to the paper.*
*\*jianhuayao@tencent.com*

**Abstract:** Artificial Intelligence (AI)-powered pathology is a revolutionary step in the world of digital pathology and shows great promise to increase both diagnosis accuracy and efficiency. However, defocus and motion blur can obscure tissue or cell characteristics hence compromising AI algorithms' accuracy and robustness in analyzing the images. In this paper, we demonstrate a deep-learning-based approach that can alleviate the defocus and motion blur of a microscopic image and output a sharper and cleaner image with retrieved fine details without prior knowledge of the blur type, blur extent and pathological stain. In this approach, a deep learning classifier is first trained to identify the image blur type. Then, two encoder-decoder networks are trained and used alone or in combination to deblur the input image. It is an end-to-end approach and introduces no corrugated artifacts as traditional blind deconvolution methods do. We test our approach on different types of pathology specimens and demonstrate great performance on image blur correction and the subsequent improvement on the diagnosis outcome of AI algorithms.

## 1. Introduction

AI is transforming digital pathology with improved efficiency and high diagnostic accuracy [1–3]. Google's AI algorithm, Lymph Node Assistant (LYNA), is better at spotting metastatic breast cancers than pathologists [4]. As more and more AI techniques are employed in digital pathology to assist pathologists in making a better diagnostic decision, the quality of images acquired by clinical microscopes and pathology slide scanners becomes ultimately critical in the process [1,5]. Defocus and motion blur are the two most common microscopic imaging artifacts that can severely degrade the digital images captured by the cameras within digital pathology instruments. They substantially increase the risk of misdiagnosis and false treatment [6]. For example, when capturing ThinPrep cytological test (TCT) images, defocus and motion blur can lead to fuzzy cell borders and distorted cell morphology, which can seriously affect cancer subtyping and grading [7].

Defocus blur can occur on images acquired by pathology slide scanners due to insufficient focusing accuracy of the autofocusing systems [8]. For clinical microscopes with cameras, defocus blur can occur due to the lack of parfocality and the optical path length differences between the eyepiece port and camera port. One prominent example is the AI-powered augmented reality (AR) microscope [5] developed by Google. In this platform, the sample slides are manually focused by pathologists when they inspect the slides though the eyepiece port. The image captured through the camera port, on the other hand, may be completely out of focus due to the optical path difference between the eyepiece port and the camera port. This optical path difference varies for different objective lenses. As a result, it is challenging to maintain the in-focus position for the camera when switching different objective lenses [9].

Similarly, motion blur results from sample movement during camera exposure. In the AR microscope platform [5], the microscope light intensity is adjusted according to the user's viewing comfort through the eyepiece. This intensity level, however, is significantly lower than what is needed for a motion-frozen camera exposure. Therefore, the acquired images are vulnerable to motion blur when the user is panning around the sample. In some of slide scanners, pulsed illumination is used to freeze the sample motion. Intensity variations of the light source needs to be calibrated over time. This approach is not suitable for AR microscope platform.

Defocus and motion blurs are often tackled separately. A smaller aperture can be used to capture an extended-depth-of-field image at the cost of spatial resolution [10]. Defocus blur can also be corrected using hardware autofocusing [11–13], computational depth-of-field extending [14,15], or deconvolution [16,17]. Some of these approaches require hardware modifications to the optical system, such as adding extra optical and electrical components like mechanical stages, liquid lenses, and phase masks. They may be incompatible with existing clinical microscopes. Deconvolution methods can correct the defocus blur without extra hardware modifications but generally suffer from ring artifacts and heavy computational load. Meanwhile, motion blur can be recovered by computational methods such as illumination engineering [18], coded exposure imaging [19], and deconvolution methods [20,21]. These methods have the same limitations stated in the defocus blur handling. Lowering the speed of stage movement or completely stop the stage to capture static images can eliminate motion blur. It, however, alters pathologists' reading routine and dramatically reduces clinical efficiency.

Recently, deep learning approaches have been demonstrated for solving traditional optical problems including super-resolution imaging [22,23], phase retrieval [24], and image deblurring [25–28], among others. Wu et al. [25] demonstrate a conditional generative adversarial neural network (cGAN) for virtually refocusing a two-dimensional fluorescence image on user-defined three-dimensional surfaces within the sample. Kupyn et al [26] achieve blind motion deblurring using cGAN. Yan et al [27] employ a general regression neural network (GRNN) to estimate the blur kernel and then apply the non-blind deconvolution method to perform image deconvolution. Zhao et al [28] propose a residual dense network(RDN) for image deblurring in various biomedical imaging problems including pathological images. These methods, however, either solve a single type of image blur (defocus blur or motion blur) or still have a traditional iterative deconvolution process in the entire image deblurring process.

Here we report a blind convolution framework for microscopic pathology images based on deep learning. Our framework can automatically deblur a single microscopic image on defocus or motion blur. We term our approach deep blind microscopic image deblurring (DBMID). In DBMID, we employ a two-phase deblurring workflow. First, we determine the blur type of an input image using a Visual Geometry Group Network (VGG16)-based classification network. We then restore the input image with the corresponding deblurring models based on very deep convolutional encoder-decoder networks [29] which are composed of multiple layers of convolution and de-convolution operators, learning end-to-end mappings from blurred images to the high-resolution ones.

We evaluate the reported DBMID using the United States air-force (USAF) 1951 resolution target and various pathological specimens, including hematoxylin and eosin (HE) stained slides, thinprep cytology test (TCT) slides, and immunohistochemistry (IHC) slides. We demonstrate how our method corrects the image blur and subsequently improves the diagnostic outcome of AI algorithms. We also investigate the generalization ability of DBMID over different microscopes.

Our approach is a cost-effective image quality enhancement solution for digitized microscopes and pathology slide scanners in a manner of introducing no extra optical or mechanical components and without hampering regular procedure of reading and scanning. We anticipate the reported method will enable better accuracy and consistency of examining pathological specimens using the microscope and digital scanners for the diagnosis of cancer and other diseases.

## 2. Methods

*2.1. The workflow of the DBMID*

The workflow of DBMID is shown in Fig. 1. We first apply a classification model to determine the blur type of the input image as one of four categories: 1) in focus, 2) defocus blur, 3) motion blur, and 4) mixed blur (defocus and motion blurs coexist). The appropriate deblurring model is then applied to restore the input image. This two-stage architecture allows us to tackle defocus and motion blur effectively in two different models.

For the blur kernel classifier, we train a VGG16 network with labeled images. For the image deblurring process, we train two separate encoder-decoder neural networks named as DL_defocus (deep learning network for defocus blur correction) and DL_motion (deep learning network for motion blur correction). Images with mixed blur are sequentially processed by DL_motion and DL_defocus. The architectures of DL_defocus and DL_motion are discussed in the next section.

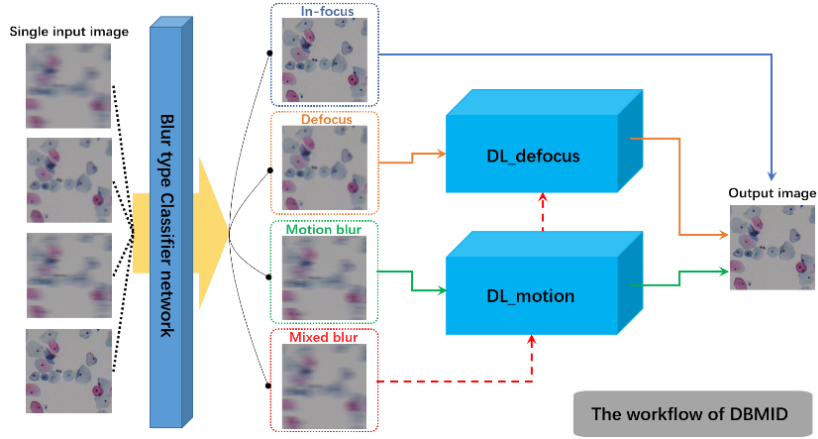

Fig. 1. The workflow of DBMID. The input images are first classified into four blur categories. For in-focus images, no operation is conducted. For defocus and motion blurred images, deblurring networks DL_defocus and DL_motion are applied accordingly. For mixed blurred images, they are sequentially processed by DL_motion and DL_defocus.

## 2.2. DL_defocus and DL_motion network architecture

Fig. 2 shows the architecture of our defocus and motion deblurring network which can be seen as an encoder-decoder network with skip connections. There are three input channels, and the input size is not restricted. The encoder module is shown at the top and the decoding part is shown below. For the numbers of layers of the encoder module and the decoder module, they should be the same, and we set it as 21 for a balance of efficiency and performance. For the size of the features of every layer, they are represented by the length and width of the block in Fig. 2. The feature size of the first layer is decided by the input image size. The downsampling and upsampling scales are both set as 2 and is implemented when using convolution or deconvolution operations with a stride of 2. The number of features of every convolution layer is represented by its thickness of the block as shown in Fig. 2 and we set them as 32, 64 and 128 from left to right both for the encoder and decoder modules empirically. The convolution kernel can be 3 or 5 and the skip connection is implemented by the sum operation. The loss function we use is the simple mean square error (MSE). For the training, we set batch size as 64 and use Adam optimizer with an initial learning rate of 1e-4. We implement the model on Tensorflow with a Tesla P40 GPU. For the input image of size 1920×1920×3, the model takes around 5 seconds per batch for the training and around 1.5 seconds for the inference.

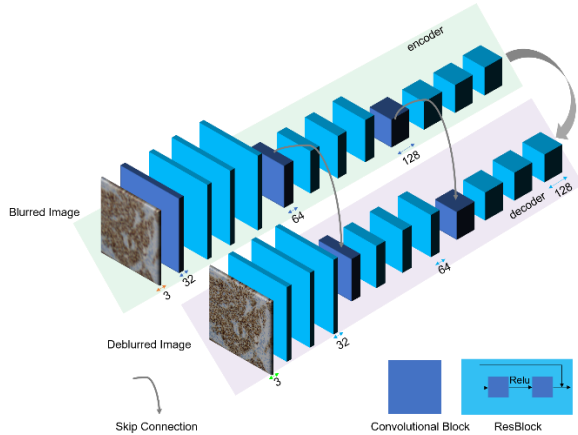

Fig 2. The encoder-decoder network structure

## 3. Experimental implementation

Our training data is acquired using an Olympus BX43 bright-field microscope with a 10X/0.3NA objective lens. We use a 4-megapixel color camera with a pixel size of 5.5 µm (Lumenera Lt425) for image acquisition. A focus-tunable liquid lens (Optotune EL-16-40-TC) is integrated into the optical system for rapid focus adjustment.

Our specimens for training and testing include HE slides, TCT slides, IHC slides, and USAF 1951 resolution target. The overview of our data is shown in Table 1, where the training and testing datasets are completely independent.

**Table 1. Training and Test Data**

|  |  | TCT | HE | IHC | USAF | Total |
|---|---|---|---|---|---|---|
| Training | In-focus | 200+200 | 200 | 200 | 0 | 800 |
|  | Defocus blur | 4000 | 4000 | 4000 | 0 | 12000 |
|  | Motion blur | 3200 | 0 | 0 | 0 | 3200 |
|  | Mixed blur | 3200 | 0 | 0 | 0 | 3200 |
| Testing | In-focus | 10+30 | 10 | 10 | 1 | 61 |
|  | Defocus blur | 200 | 200 | 200 | 20 | 620 |
|  | Motion blur | 160+160 | 0 | 0 | 0 | 320 |
|  | Mixed blur | 320 | 0 | 0 | 0 | 320 |
| **Total** |  | **11680** | **4410** | **4410** | **21** | **20521** |

For the training of the DL_defocus network, we use 200 in-focus images and 4000 defocused images at 20 focal planes for HE, TCT, and IHC specimens. The input sample size is 1920×1920, and all the image pairs are registered to the in-focus images using ImageJ plugin 'StackReg' for defocus images. The testing of defocus blur contains 10 in-focus images and 200 defocused images for the three kinds of histology slides and 1 in-focus image and 20 defocused images for the USAF sample.

For the DL_motion network, the in-focus images are used to simulate motion-blurred images using a uniform blur kernel of size 5 to 40 pixels with a step size of 5 in both horizontal and vertical directions, and we synthesize 3200 and 160 blurred images for the TCT samples as the training and testing dataset, respectively.

For the blur type classification network, extra 200 and 30 TCT in-focus images are added for the training and testing process, respectively. We also synthesize 3200 mixed blurred images for training, 320 mixed blurred images and extra 160 motion blurred images for testing. We have made our code, dataset, the data acquisition details, other methods and evaluation metrics available for the broad research community, as we show in Code 1 (Ref. [30]).

## 4. Experimental results

*4.1 Classification performance of blur type*

The classification accuracy on the testing is 99.77%, and the confusion matrix is shown in Table 2. Only 2 in-focus images are misclassified as with defocus blur.

**Table 2. The Confusion Matrix for the Blur Type Classification**

|  |  | Actual Class | | | |
| --- | --- | --- | --- | --- | --- |
|  |  | In-focus | Defocus | Motion blur | Mixed blur |
|  | In-focus | 38 | 0 | 0 | 0 |
| Predicted | Defocus blur | 2 | 200 | 0 | 0 |
| Class | Motion blur | 0 | 0 | 320 | 0 |
|  | Mixed blur | 0 | 0 | 0 | 320 |

*4.2 Defocus deblurring of the USAF resolution target*

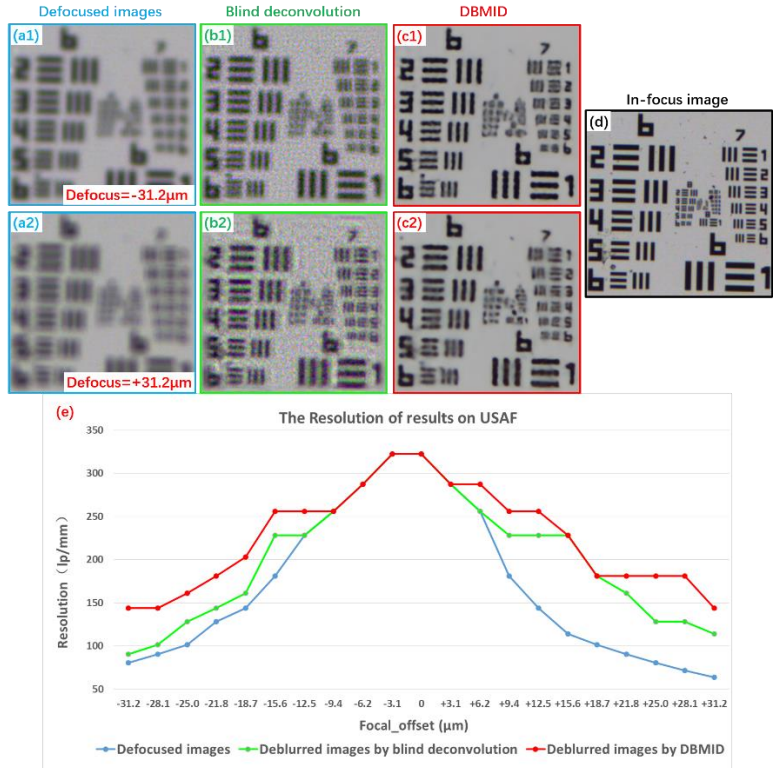

Fig. 3. Image quality improvement by traditional blind deconvolution method and our DBMID on the USAF resolution target. (a1)-(a2) are the defocus blurred images with focal offsets of -31.2 µm and +31.2 µm. (b1)-(b2) are the deblurred results by the traditional blind deconvolution method. (c1)-(c2) are the deblurred results by our DBMID. (d) is the in-focus reference image. (e) Plots of the resolution values over the focal offset extending from -31.2 µm to +31.2 µm.

Defocus blur reduces the sharpness and contrast of the image, thus compromising the post-processing and diagnostic results. We first evaluate our model's defocus deblurring capability using the USAF resolution target. In this experiment, the reported method is compared with the traditional blind deconvolution approach which requires no prior information of the blur kernel.

Fig. 3 (a1) and (a2) show the defocused images captured at defocus distances of -31.2μm and +31.2μm, respectively. Here '+' denotes the image plane above the in-focus position (closer to the objective) and '-' denotes the image plane below the in-focus position (further away from the objective). From Fig. 3 (a1) and (a2), we can see that the axial defocused images lack fine details compared with the in-focus images in Fig. 3(d). Both traditional deconvolution method and the reported DBMID can enhance the resolution and contrast in Fig. 3(b), 3(c), and the quantitative results in 3(e). However, color and granular artifacts are introduced to the image background for the deconvolution method. These artifacts are a result of noise amplification and imperfect point spread function (PSF) estimation. On the other hand, DBMID introduces almost no observable artifacts to the deblurred images and correctly focuses the images with different defocus extents. It should be noted that the USAF target images have not been included in the training set, thus demonstrating the generalizability of the reported DBMID.

*4.3 Defocus deblurring for microscopic pathology images*

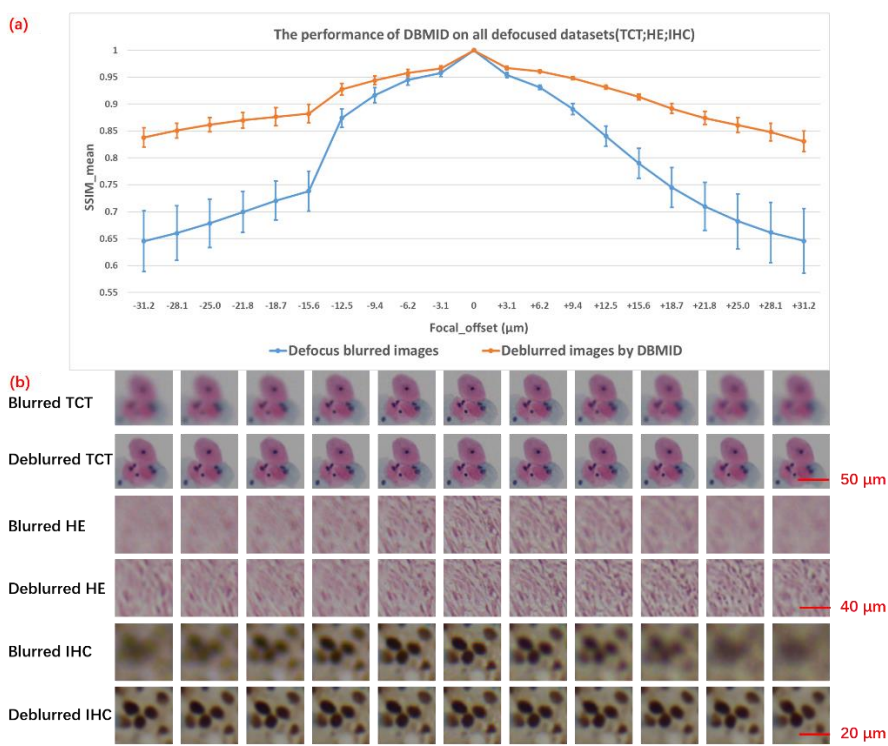

Fig. 4. Defocus deblurring for microscopic pathology specimens. (a) Quantitative analysis of defocus deblurring capability of the DBMID model via the SSIM values. The blue curve represents the SSIM value of the original defocus blurred images. The orange curve represents the SSIM value of the deblurred images by DBMID. (b) Examples of defocus blurred and deblurred images of TCT, HE and IHC over the focal axis direction.

To make a thorough evaluation of the deblurring capability of DBMID, we also test different types of pathology specimens (TCT, HE and IHC) in our experiments.

In Fig. 4 (a), we plot the statistical SSIM values (structural similarity index) between the focused images and the restored images at different defocus planes. The results are calculated and averaged over all the datasets. We can see that both the mean and the standard deviation of SSIM values of the deblurred images are statistically improved over those of the defocused images, for a very wide range along the focal axis. Fig. 4(b) also confirm this with examples of defocused and deblurred images using DBMID. The depth of field is about 9.8 μm in the original images, whereas the depth of focus is increased to 31.2 μm or more after the deblurring operation, which is sufficient for most microscopy applications. From the results in Fig. 4(b), we can also see that the DBMID can automatically adjust the deblurring power for both the severely defocused and the mildly defocused images, recovering missing textures while not introducing artifacts. This is the advantage of our deep learning-based method which can extract, combine and analyze features automatically.

As seen in Fig. 4(b), the appearances of different types of microscopic pathology images are vastly different. For DBMID, it is trained with all types of pathology images to improve the model generalizability. As a comparison, we also train separate models using individual TCT, HE and IHC dataset, named as DL_defocus_one. These individual models are tested on their respective dataset and the results are summarized in Table 3. The model trained with all types of images (DL_defocus_all) achieves comparable or better results than those trained using only one image type (DL_defocus_one).

**Table 3. Results of Model Generalizability**

| SSIM | Defocus blur | DL_defocus_one | DL_defocus_all |
|---|---|---|---|
| **TCT dataset** | 0.9007±0.0288 | 0.9477±0.0100 | 0.9458±0.0106 |
| **HE dataset** | 0.6679±0.1898 | 0.8126±0.1045 | 0.8281±0.0941 |
| **IHC dataset** | 0.7846±0.1431 | 0.9280±0.0379 | 0.9261±0.0384 |
| **All datasets** | 0.7844±0.1677 | 0.8961±0.0878 | 0.9000±0.0783 |

*4.4 Motion deblurring via DBMID*

In this section, we demonstrate the ability to recover motion-blurred images using the reported DBMID. Motion blur can seriously degrade the image quality, caused by the movement of the image being recorded during the camera exposure time. We use TCT images with different motion blur levels and directions in the experiment. Two TCT images with a 20-pixel motion blur in the horizontal and vertical directions are shown in Fig. 5(a1) and (a2), respectively. Figure 5(b) shows the corresponding recovered images using the conventional blind deconvolution method and our DBMID results are shown in Fig. 5(c). To quantify the motion deblurring capability of DBMID, we plot the averaged lateral full-width-half-maximum (FWHM) values of 60 individual immune cells in Fig. 5(e) and (f) for the blurred and deblurred images with different motion blur levels and directions. These 60 immune cells are selected from 10 TCT image views with6 from each in-focus image. FWHM values of in-focus and the corresponding blurred images are calculated from the same cropped area. The FWHM measurement is similar to point spread function analysis in fluorescence imaging using nanobeads [31]. DBMID shows consistently good performance as the motion blur level varies. The FWHM gap between the blurred and deblurred images increases as the motion blur level increases. On the other hand, the deblurred results by the DBMID method match well with the in-focus reference image. As for the blind deconvolution method, it shows comparable blind deblurring capability but suffers from ring artifacts.

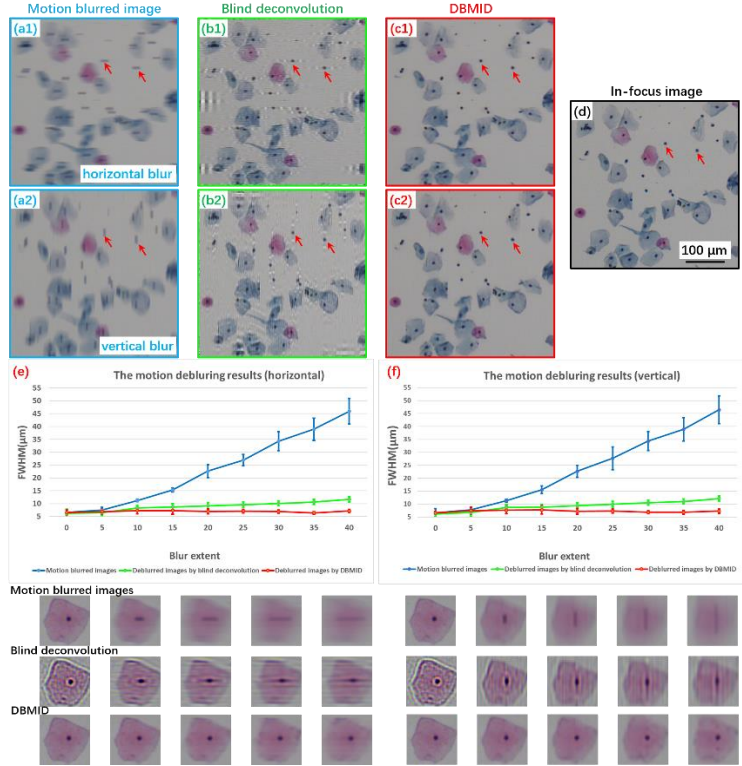

Fig. 5. Motion blur correction by blind deconvolution and our DBMID on TCT images. Horizontal (a1) and Vertical (a2) motion-blurred TCT images with a 20-pixel motion blur. (b1) and (b2) are blind deconvolution results. (c1) and (c2) are our DBMID deblurred images. (d) is the in-focus reference image. (e) shows the lateral full-width-half-maximum (FWHM) values of 60 individual immune cells (red arrows pointed in (a)-(d)) for horizontally motion-blurred and deblurred images. (f) shows the lateral FWHM values of 60 individual immune cells (red arrows pointed in (a)-(d)) for vertically motion-blurred and deblurred images. A few examples of various motion blur extents are shown at the bottom.

### 4.5 Mixed deblurring via DBMID

Both defocus and motion blur may occur simultaneously (mixed blur) in microscopic images when the specimen is out-of-focus while the user moves the slide. In this section, we show that DBMID can deal with the mixed blur by successively applying DL_motion and DL_defocus networks. A cropped image of a TCT sample is shown in Fig. 6(a), which contains both defocus blur ($z = +15.6$ µm) and a 10-pixel motion blur in the vertical direction. Figure 6(b) and 6(c) show the recovered image using a single deblurring network of DL_defocus and DL_motion, respectively. Figure 6(d) shows the DBMID deblurred results of Fig. 6(a), which outperforms any single deblurring results shown in Fig. 6(b) and 6(c). In Fig. 6(d), we can find that the smear of the nuclei is eliminated, and the cell borders are much clearer. Fig. 6(f)-(i) shows the quantitative results of SSIM heatmaps of various focal offsets and motion blur extents. DBMID for mixed deblurring clearly outperforms individual deblurring network.

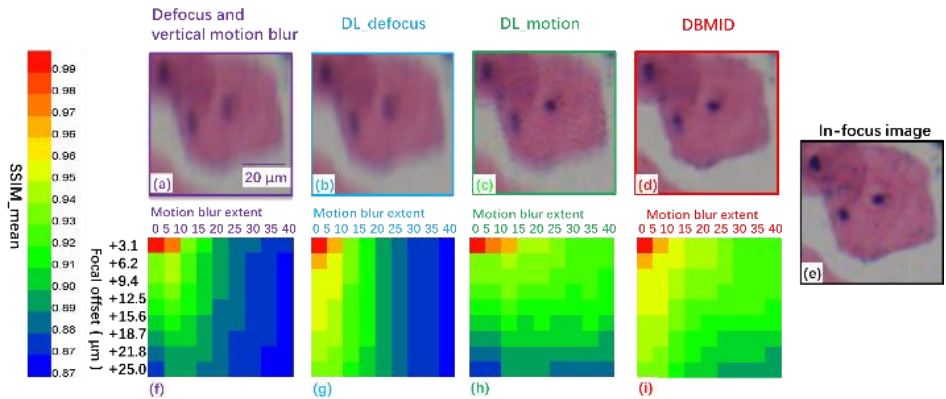

Fig. 6. Mixed deblurring using DBMID. (a) An image with both defocus and motion blurs. (b) The DL_defocus deblurred result of (a). (c) The DL_motion deblurred result of (a). (d) The DBMID deblurred result of (a). (e) The in-focus image. (f)-(i) The SSIM heat maps of various focal offsets and motion blur extents for the mixed blurred image, the deblurred images by DL_defocus, by DL_motion, and by DBMID, respectively.

*4.6 Influence of DBMID on detecting cancerous cervical cells*

TCT screening test plays an essential role in identifying precancerous lesions that are caused by human papillomavirus (HPV) and cervical cancers at an early stage. Detecting the cancerous cervical cells from thousands of candidates on a TCT slide is the prerequisites for the diagnosis. An AI algorithm has been developed to detect cancerous cervical cells on TCT images acquired by slide scanner to help improve pathologists' efficiency. Here we evaluate the performance of the AI algorithm on in-focus, defocused and deblurred TCT images. Some sample images are shown in Fig. 7.

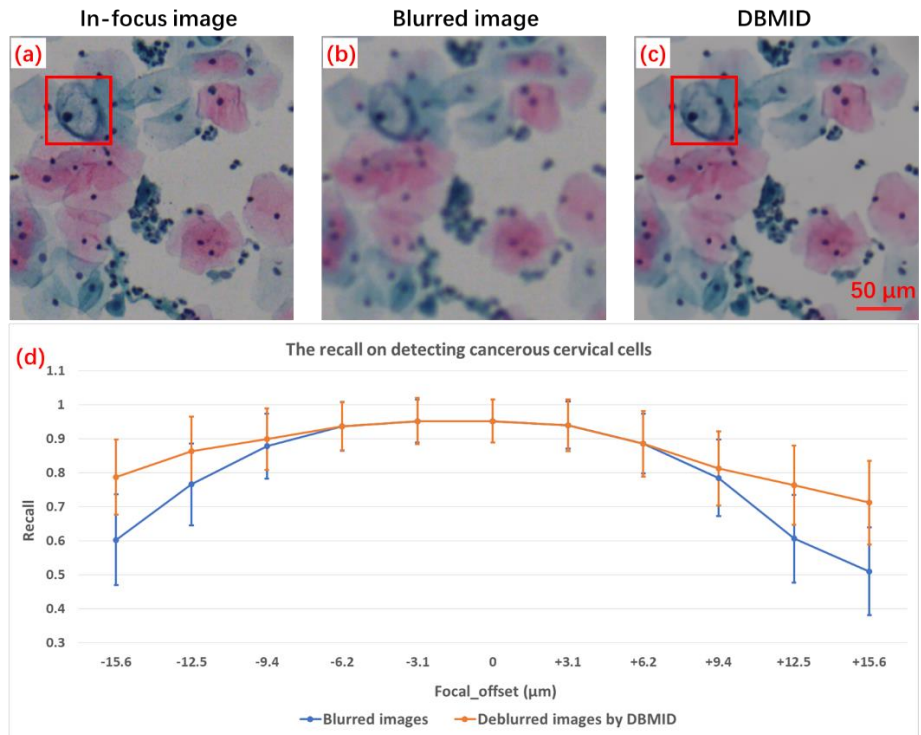

Fig.7. AI-powered cancerous cervical cell detection on blurred and deblurred TCT images. (a) An in-focus TCT image. (b) The corresponding blurred TCT images. (c) The deblurred results using DBMID on (b). The red rectangles indicate

the detected positive cervical cells by applying the AI algorithm. (d) The recall score improvement using the deblurred images.

Defocus blur smooths out the cell morphological and texture details that differentiate the cancerous cells from the normal cells. For instance, a cancerous cervical cell inside the red box shown in Fig. 7(a) cannot be recognized using the blurred image of Fig. 7(b). DBMID rescues most of the morphological characteristics and intracellular details of the cells, making the cancer cell distinguishable by the detection algorithm (Fig. 7(c)). The recall improvement in Fig. 7(d) also confirms this especially when the image is severely blurred.

*4.7 Influence of DBMID on IHC quantification*

In this section, we evaluate how our method improves the performance of an AI algorithm conducting progesterone receptor (PR) expression quantification on breast immunohistochemical (IHC) images. The AI algorithm first detects the nuclei within tumor regions, and then computes the percentage of positively stained tumor nuclei among all nuclei [32,33], denoted by PR Rate. With many touching cells in the images, blurred images affect the algorithm's capability of separating cells and thus obtaining the accurate PR Rate.

To comprehensively assess the performance of the AI algorithm on the blurred and deblurred IHC images, eight views are chosen within a PR stained pathology slide. For each location, the image stacks are acquired at defocus planes ranging from -15.6 μm to +15.6 μm from the in-focus position. In Fig. 8, we evaluate the AI performance of cancer nucleus detection, cancer region segmentation and PR rates on the in-focus, defocused and deblurred images.

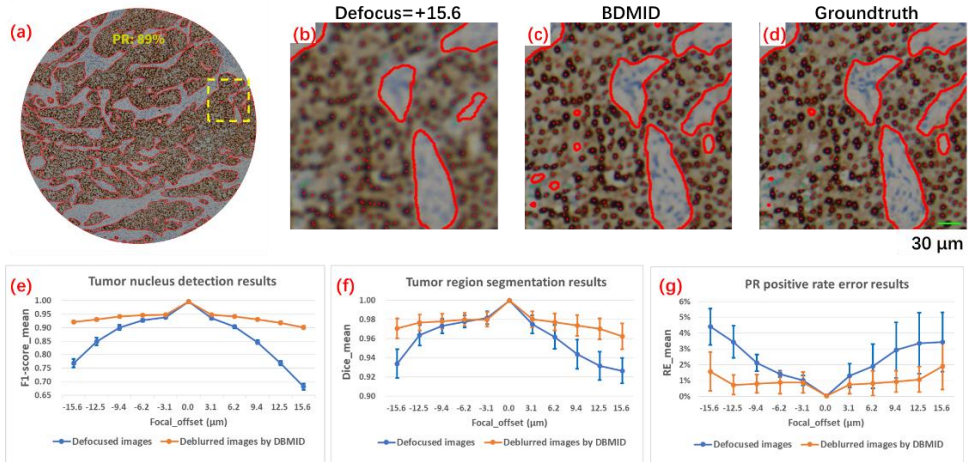

Fig. 8. Influence of DBMID on IHC quantification. (a) An example image with PR=89%. (b-d) The cropped and zoomed-up patches of tumor nucleus detection and region segmentation. (b) The AI result on a defocused image (z = +15.6 µm). (c) The AI result on the BDMID deblurred image. (d) The ground truth result. (e) Tumor nucleus detection results (measured by F1-score_mean) comparison between defocused images and DBMID deblurred images. (f) Tumor region segmentation results (measured by Dice_mean) comparison between defocused images and DBMID deblurred images. (g) Tumor nuclei positive rate results (measured by RE_mean) comparison between defocused images and DBMID deblurred images.

The detected nuclei (namely the red points for positively stained nuclei and green points for negatively stained nuclei) and tumor region contours (namely the red curves) are shown on the zoomed views in Fig. 8(b) and (c). It is illustrated that considerably blurred nuclei (e.g., in Fig. 8(b)), even originally stained clearly in Fig. 8(a), are subject to detection misses, which can be correctly identified on the DBMID deblurred images. From a quantitative perspective, the DBMID deblurred images could lead to considerably better performance compared to the corresponding defocused images along the focal axis, especially when the defocus distance is relatively larger as shown in Fig. 8(e)-(g). For instance, the F1-scores corresponding to the defocus images at z>+11 µm and z<-14 µm in Fig. 8(e) are all below 0.80. After deblurring, they are all improved to higher than 0.92. Figure 8(f) shows the tumor region segmentation result that is relatively less affected by image blurriness. There is still a slight improvement in the results of DBMID deblurred images compared with that of the blurred images. Via a visual comparison of the highlighted contours in Fig. 8(b)-(d), we can conclude that better image quality thanks to the deblurring operation could help differentiate even small non-tumor regions from the neighboring tumor regions. Most importantly, Fig. 8(g) also demonstrates that the deblurred images effectively reduce the error in PR-positive rate to less than 1% for the defocus levels from -12.5 µm to +12.5 µm. Defocused images suffer from up to ~3.5% error rate at the same defocus range.

Consequently, the DBMID method greatly assists the IHC algorithm in fulfilling the accuracy requirements of PR quantification, especially when the rate is around any clinical cutoff values [33]. It is of significance to segregate patients who are likely to attain clinical benefits from those who will not. Based on the comprehensive evaluation with the above three metrics, it is beneficial to preprocess defocused IHC images by the DBMID for an accurate and reliable PR-positive rate.

*4.8 Generalization of DBMID over different microscopes*

The above-mentioned experiments are done on Olympus BX43 with a 10X/NA0.3 objective lens. To demonstrate the generalization ability of DBMID, we evaluate the performance on a Sunny Optics RX50 microscope with a 10X/NA0.25 objective lens as external validation. For a fair comparison, we acquire the same set of IHC images on Sunny Optics microscope as those in the experiments using the Olympus microscope.

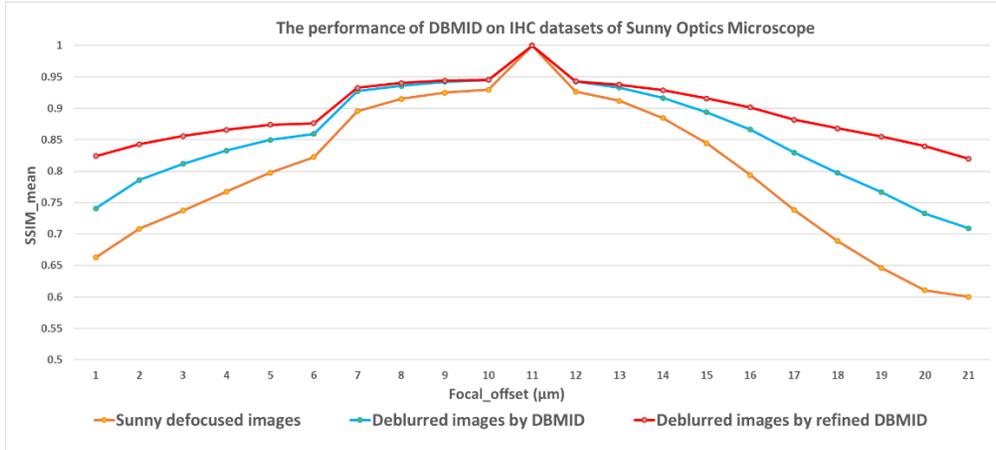

Fig. 9. The performance of DBMID on IHC datasets acquired by Sunny Optics microscope

We conduct two experiments to validate the generalizability. First, we apply the original DBMID model trained on Olympus images on the defocused Sunny images (blue line on Fig 9). Second, we fine-tune the original DBMID model using Sunny training images accounted for one-tenth of the orignal training set and apply it on the Sunny test set. (red line on Fig 9). As Fig. 9 shows, DBMID trained on Olympus images can deblur the images acquired by Sunny Optics microscope. The fine-tuned model further improves the performance, especially for blurred images with large focal offset. These experiments demonstrates the generalization ability of DBMID.

5. **Discussion**

Defocus blur and motion blur are the two most common blurs in microscopic images. They can severely reduce the image contrast and sharpness, hampering the image-based diagnostic process. Our deep learning approach, termed DBMID, permits the recovery of the sharp image without prior knowledge of the blur type, blur extent and histological stain type.

We have demonstrated how DBMID can remove the defocus blur and extend the depth of field of the imaging system. From Figs. 3 and 4, we can observe significant image quality improvement on the USAF target, TCT, HE and IHC samples within a defocus range from -31.2 µm to +31.2 µm. We note that, this defocus range is just the practical defocus distance in a regular microscope setting but not the deblurring limit of DBMID. We can train DBMID with deeper defocused data to achieve a better depth of field enhancement. On the other hand, image blur level is mainly influenced by the optical system parameters such as numerical aperture, optical system aberrations, defocus distance, etc. DBMID learns the optical system parameters and can deblur the USAF target 1951 which has not been used in the training data. We demonstrate the generalizability of our model by showing its deblurring effect on various kinds of pathology images and other microscopes. The defocus deblurring capability of DBMID proves it a cost-effective alternative solution to hardware autofocusing systems for microscopy.

We also validate the motion deblurring capability of DBMID in this work. The DBMID deblurred results match well with the ground truth images. The FWHM values of the deblurred small immune cells retain the same scale as those of the in-focus images. The presented results further confirm that the motion blur has been effectively removed from the blurry images. The motion-deblurring capability of DBMID provides a remedy for previously collected motion-blurred images. It also reduces the hardware requirements for optical systems such as the light source intensity and stage movement speed.

Benefiting from the special structure of DBMID, it can handle different blur types without prior knowledge. The DBMID framework first uses a VGG model to classify the blur type of the input image. It then feeds the blurred image to the appropriate deblurring model such as DL_defocus, DL_motion or both. Table 2 shows that the blur type classification can be accurately performed using the reported network. We achieved 99.77% accuracy solely relying on spatial information of the input image. One can further improve the classification network by considering both spatial and Fourier spectrum inputs [34]. For example, we can find fringe patterns in Fourier magnitude of motion blurred images, which does not exist in non-motion blurred images. Besides, the frequency and direction of the fringe patterns in the motion blurred image are related to the motion blur level and direction. Deblurring mixed blur is a challenging task for conventional blind deconvolution. To address this problem, we propose to concatenate DL_motion and DL_defocus and remove motion blur and defocus blur in order.

We have also demonstrated how DBMID can assist AI algorithms in getting better results on pathology analysis. DBMID can remove the blur and retrieve the fine details which are lost in the blurred images. For example, the DBMID method assists the IHC algorithm in fulfilling the accuracy requirements of PR quantification, especially when the PR Rate is around any clinical cutoff values [33]. It is beneficial to preprocess the defocused IHC images using the DBMID for a reliable evaluation. As AI is an emerging tool for pathology diagnosis, DBMID has the potential to improve the image quality and contribute to various applications not limited to the cancerous cervical cell detection and IHC sample analysis discussed in this work.

Classic image deblurring approach requires explicit analytical modeling of the point spread function of the optical system and noise level. Otherwise, it is a challenging ill-posed inverse problem with unknown kernel information. The iterative deconvolution process often requires regularization techniques to suppress undesirable artifacts and stabilize the restoration process. However, it is hard to find a perfect regularizer to tackle both image resolution enhancement and artifacts suppression. The data-driven approach as the DBMID we proposed, can automatically learn the end-to-end image recover process from the blurred-focused image pairs without the need for analytical modeling and handcrafted transformations. DBMID provides sharp and clean deblurred results that don't have disturbing artifacts that commonly exist due in the deconvolution process. This is the unparalleled advantage of deep neural networks compared to classic analytical methods.

One inherent limitation of DBMID is that it takes considerable time for image deblurring and thus not an ideal solution for real-time imaging applications. Take a full resolution 1920*1920 color image as an example, it takes about 1.5s on average by DBMID to finish image deblurring using a Tesla P40 GPU. However, DBMID still outperforms the blind deconvolution method which takes around 13.5 s on average in this case. Further research about speeding up the deblurring network is under way. Another limitation of DBMID is that currently it only handles two most common types of blurring in optical microscopic systems: defocus blur and linear motion blur. We would need to fine-tune the model for other less common types of optical microscopic blurs such as high order aberrations, chromatic aberrations, jitter and zig-zag motion blur.

We believe the presented framework can find broad applications in other biomedical imaging, such as whole slide scanners and regular cell inspection. Beyond microscopic imaging, it may also provide new insights for photographic deblurring innovations.

**Disclosures**

The authors declare no conflicts of interest.